\documentclass[twocolumn]{elsart}
\usepackage{amssymb}
\usepackage{epsfig}

\begin{document}

\begin{frontmatter}

\title{Frustration phenomena in Josephson junction arrays on a dice lattice}
\author{M. Tesei},
\author{R. Th\'{e}ron},
\author{and P. Martinoli}
\address{Institut de Physique, Universit\'{e} de Neuch\^{a}tel, 2000 Neuch\^{a}tel, Switzerland}

\begin{abstract}
AC magnetoimpedance measurements performed on proximity-effect
coupled Josephson junction arrays on a dice lattice reveal
unconventional behaviour resulting from the interplay between the
frustration $f$ created by the applied magnetic field and the
particular geometry of the system. While the inverse
magnetoinductance exhibits prominent peaks at $f=1/3$ and at
$f=1/6$ (and weaker structures at $f=1/9,2/9,1/12,...$) reflecting
vortex states with a high degree of superconducting phase
coherence, the deep minimum at $f=1/2$ points to a state in which
the phase coherence is strongly suppressed. These observations are
discussed at the light of recent theoretical work in which the
concept of accidental degeneracy plays a central role.

\end{abstract}


\end{frontmatter}
% ----------------------------------------------------------------

\begin{figure}[h]
 \begin{center}
 \includegraphics[width=6cm]{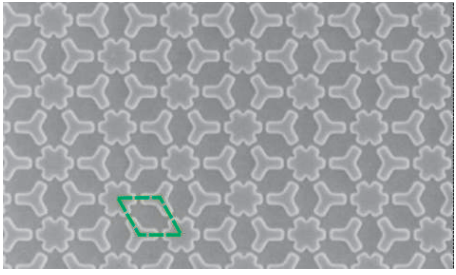}
 \caption{\label{sample}SEM picture of a portion of a JJA on a dice lattice.
 The superconducting lead islands are Josephson-coupled by an underlying Cu layer
 (dark ground plane). The elementary cell is rhombic in shape (dashed line)
 and has a side $a=8\mu m$.}
 \end{center}
\end{figure}

\begin{figure}[h]
 \begin{center}
 \hspace*{-0.5cm}\includegraphics[width=7.5cm]{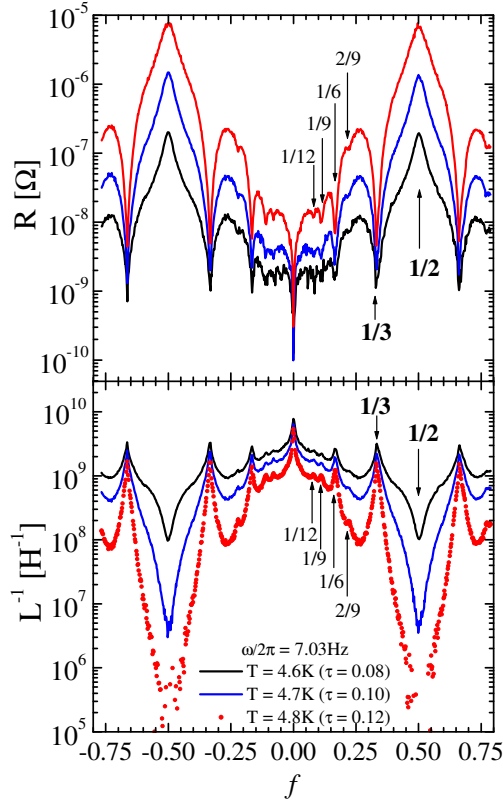}
 \vspace{-1cm}\caption{\label{RLinv(f)}Magnetoresistance isotherms
 (top) and inverse magnetoinductance isotherms (bottom) for a JJA on a
 dice lattice measured at an excitation frequency $\omega/2\pi=7.03$Hz.}
 \end{center}
\end{figure}

Two-dimensional Josephson junction arrays (JJAs) exposed to a
transverse magnetic field provide the opportunity to study the
influence of a tunable level of frustration in systems with a
variety of geometries ranging from periodic to random structures,
including quasiperiodic and fractal lattices
\cite{ReviewMartinoli}. Such systems are usually considered as a
physical realization of the frustrated classical XY model
\cite{TeitelJayaPRB27}, where the degree of frustration is
governed by a parameter $f$ expressing the magnetic flux threading
an elementary cell, $\Phi_{cell}$, of the array in units of the
superconducting flux quantum $\phi_{0}$ :
$f=\Phi_{cell}/\phi_{0}$.\\
In this contribution we study the interplay of frustration and
geometry in proximity-effect coupled JJAs on an unconventional
lattice, the dice lattice shown in Fig.\ref{sample}, by measuring
the complex sheet impedance $Z(T,\omega,f)=R+i\omega L$ of the
system with a SQUID-operated two-coil mutual inductance technique
\cite{Jeanneret_2coils}.
The inverse sheet inductance $L^{-1}=Im[Z]/\omega$, which is
proportional to the areal superfluid density, measures the degree
of superconducting phase coherence in the sample, and the sheet
resistance $R$ reflects dissipative processes. Inverse
magnetoinductance $L^{-1}(f)$ and magnetoresistance $R(f)$
isotherms are shown in Fig.\ref{RLinv(f)}, $\tau=k_{B}T/J(T)$
being the relevant reduced temperature expressed in terms of the
Josephson coupling energy.

The prominent peaks appearing in $L^{-1}(f)$ at $f=1/3$ and
$f=1/6$ point to vortex states with a high degree of
superconducting phase coherence, and are robust against thermal
fluctuations: for instance, the height of the structure in
$L^{-1}(f)$ at $f=1/3$ changes only by a factor two in the
temperature range covered by the data of Fig.\ref{RLinv(f)}.
Weaker peaks in $L^{-1}(f)$ at $f=1/9,2/9,1/12$ are also a
manifestation of vortex states with an appreciable degree of phase
coherence, however more vulnerable to thermal fluctuations. This
is easily understood if one realizes that the corresponding
periodic ground states consist of unit cells larger than those for
$f=1/3$ and $f=1/6$, thereby implying superconducting phase
coherence to extend at larger length scales and, therefore, to be
less robust against thermal fluctuations.\\

In striking contrast with the behaviour at $f=1/3,1/6,1/9...$, the
deep minimum in $L^{-1}(f)$ at full frustration ($f=1/2$)
indicates a strong suppression of the phase coherence, hence a
state quite vulnerable to thermal fluctuations: in the temperature
range of the data shown in Fig.\ref{RLinv(f)}, the strength of the
dip changes by at least two orders of magnitude. As expected, the
$R(f)$ curves show absolute maxima at $f=1/2$ and local minima at
$f=1/3,2/9,1/6,1/9,1/12...$,
corresponding, respectively, to dips and peaks in $L^{-1}(f)$.\\

In order to understand this unusual behaviour, we compare the
occurrence of ($i$) vortex ordering and ($ii$) superconducting
phase coherence at $f=1/2$ and $f=1/3$, the phase transitions
($i$) and ($ii$) being driven by different topological
excitations.\\

\begin{figure}[h]
 \begin{center}
 \hspace*{-0.5cm}\includegraphics[width=7.5cm]{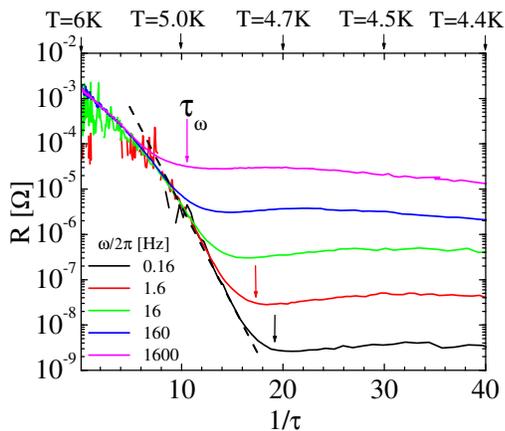}
 \vspace{-1cm}\caption{\label{R(T)}Sheet resistance versus
 inverse reduced temperature measured over a wide frequency range
 at full frustration ($f=1/2$).}
 \end{center}
\end{figure}

Allowing for the formation of zero-energy domain walls (DWs), the
ground state (GS) at full frustration ($f=1/2$) exhibits a well
developed accidental (i.e. not related to symmetry) degeneracy
\cite{Korshunov_dicedemi_GS}. Within the framework of the
uniformly frustrated XY model, this degeneracy is removed by
considering the free-energy difference (due to small-amplitude
thermal fluctuations) between ground states corresponding to
different periodic vortex patterns \cite{Korshunov_dicedemi}. This
\textit{order-from-disorder} mechanism is so weak that  a phase
transition to a disordered vortex pattern due to the proliferation
of DWs (with an almost concomitant suppression of superconducting
phase coherence resulting from the unbinding of fractional
vortices \cite{Korshunov_dicedemi}) is expected to occur only at
very low temperatures ($\tau_{c}\approx 0.01$) and in samples of
macroscopic size ($L\gg 10^{5}a$) \cite{Korshunov_dicedemi}.
However, in our proximity-effect coupled arrays, magnetic effects
associated with the interactions of screening currents become more
and more relevant with decreasing temperature, thereby providing a
more efficient mechanism to lift the accidental degeneracy of the
ground state \cite{Korshunov_dicedemi}. In principle, therefore,
at $f=1/2$ the disordering of the low temperature periodic
vortex-pattern (which simultaneously drives the suppression of
superconducting phase coherence) should set in above
$\tau_{c}\approx 0.01$. However, no evidence for a genuine phase
transition, as it would be reflected in the appearance, at
$f=1/2$, of a \textit{peak} in $L^{-1}(f)$ and a \textit{dip} in
$R(f)$, is found in the data of Fig.\ref{RLinv(f)}. A possible
explanation is that the high-temperature liquid-like vortex state
resulting from the proliferation of DWs progressively evolves,
with decreasing temperature, to a \textit{dynamically frozen}
vortex liquid characterized by a glass-like dynamics
\cite{Calame_crossover}. In this scenario, a regime crossover,
rather than a genuine phase transition, is expected, the crossover
temperature $\tau_{\omega}$ depending on the time scale $1/\omega$
of our measurements. The Arrhenius plot of Fig.\ref{R(T)} is
clearly consistent with this interpretation. The crossover
temperature $\tau_{\omega}$ separates the frequency-independent
exponential behaviour of $R(\tau)$ characteristic of a vortex
liquid at $\tau>\tau_{\omega}$ from the frequency-dependent
frozen-liquid regime at $\tau<\tau_{\omega}$
\cite{Calame_crossover}.

It should be noticed that our interpretation of the
\textit{$f=1/2$-anomaly} in JJAs on a dice lattice relies on the
idea of accidental degeneracy and is therefore fundamentally
different from the Ginzburg-Landau mean-field treatment
\cite{Vidal} invoked in Ref.\cite{Abilio} to explain the
depression of $T_{c}$ and of the critical current in fully
frustrated superconducting wire networks on the same lattice. The
existence of an analogous accidental degeneracy in such systems
has been demonstrated in Ref.\cite{Korshunov_Doucot_diceFFWN}.\\

Within the framework of the uniformly frustrated XY model, at
$f=1/3$ the accidental degeneracy of the GS is so well developed
that the vortex pattern is predicted \cite{Korshunov_dicetiers} to
remain disordered at very low temperature. Nevertheless, at low
temperatures the system is characterized by a non vanishing
helicity modulus, a behaviour consistent with the presence of a
pronounced \textit{superfluid peak} in $L^{-1}(f)$ at $f=1/3$ (see
Fig.\ref{RLinv(f)}).\\
At low temperature, superconducting phase coherence is stabilized
by pairs of bound fractional vortices and antivortices carrying a
half-integer topological charge \cite{Korshunov_dicetiers}. Thus,
the superconducting-to-normal phase transition at $f=1/3$ is
expected to be completely analogous to the
Berezinskii-Kosterlitz-Thouless (BKT) transition of the
unfrustrated ($f=0$) system, and will be discussed in detail in a
subsequent paper.\\
A uniformly frustrated JJA on a dice lattice at $f=1/3$ is
therefore a unique example of 2D superconductor where
quasi-long-range phase coherence coexists, at any temperature,
below the BKT transition, with a disordered vortex pattern due to
the proliferation of domain walls.\\

We would like to thank S.E. Korshunov for several illuminating
discussions. This work was supported by the Swiss National Science
Foundation, and the National Center of Competence in Research
(NCCR) "Materials with Novel Electronic Properties" (MaNEP).

\end{document}